\documentclass[notoc,12pt]{JHEP3}

\usepackage{amsmath,amssymb,euscript,graphicx,cite}

\def\N{{\cal N}}
\def\ms{{\mathfrak M}}
\def\Q{{\mathfrak Q}}
\def\P{{\bf P}}
\def\Z{{\bf Z}}

\def\X{{\bf X}}
\def\J{{\EuScript J}}
\def\Y{{\bf Y}}

\newcommand{\EQ}[1]{\begin{equation} #1 \end{equation}}

\title{The Curve
of Compactified 6D Gauge Theories and Integrable Systems}
\author{Harry W. Braden\\ School of Mathematics, University of
  Edinburgh, Edinburgh, EH9 3JZ, UK \\ E-mail: \email{hwb@ed.ac.uk}}
\author{Timothy J. Hollowood\\ Department of Physics,
University of Wales Swansea, Swansea, SA2 8PP, UK\\ E-mail:
\email{t.hollowood@swan.ac.uk}} \abstract{We analyze the
Seiberg-Witten curve of the six-dimensional $\N=(1,1)$ gauge
theory compactified on a torus to four dimensions. The effective
theory in four dimensions is a deformation of the $\N=2^*$ theory.
The curve is naturally holomorphically embedding in a slanted
four-torus---actually an abelian surface---a
set-up that is natural in Witten's M-theory
construction of $\N=2$ theories. We then show that the curve can
be interpreted as the spectral curve of an integrable system which
generalizes the $N$-body elliptic Calogero-Moser and
Ruijsenaars-Schneider systems in that both the positions and
momenta take values in compact spaces. It turns out that the
resulting system is not simply doubly elliptic, rather the
positions and momenta, as two-vectors, take values in the ambient
abelian surface. We analyze the two-body system in some detail. The
system we uncover provides a concrete realization of a
Beauville-Mukai system based on an abelian surface rather than
a K3 surface.}

\preprint{SWAT-\\ EMPG-}

\begin{document}

\section{Introduction}

The Coulomb branch of four-dimensional gauge theories with $\N=2$
supersymmetry is described at low energy by Seiberg-Witten
theory \cite{SW94}. In essence, there is a complex curve, or
Riemann surface, whose moduli vary across the Coulomb branch, and
whose period matrix determines the couplings of the low energy
effective action. Theories with eight
supercharges can be defined in higher dimensions, specifically in
five and six dimensions. When compactified on a circle and torus, respectively,
where we keep the scale of the compactification space finite, this
gives rise to a four-dimensional gauge theory with $\N=2$
symmetry. Hence, we expect that there is a Seiberg-Witten curve
which controls the couplings in the low energy effective action
which now depends on the details of the compactification space. In
particular, we will consider the $U(N)$ gauge theory with a
massive adjoint hypermultiplet, sometimes called the $\N=2^*$
theory.

The $\N=2^*$ theory can be lifted to
six dimensions where it is has $\N=(1,1)$ supersymmetry, although
strictly in six dimensions the hypermultiplet is massless.
Note that the six-dimensional gauge
theory is non-renormalizable, but it may be given an ultra-violet
completion as a little string theory: the theory describing five
branes in Type II string theory. Once compactified on a torus
the resulting theory is an $\N=2$  $U(N)$
gauge theory in four dimensions
with a complexified coupling $\tau$ given by $g_6^{-2}$
times the complexified K\"ahler class of the torus. The
effective four-dimensional theory will also depend on the
complex structure of the torus $\rho$ and, in
addition, one can include a certain twist in the R-symmetry group
which gives a mass $m$ to the hypermultiplet in four dimensions.
By a twist, we
mean that there are non-trivial holonomies in a $U(1)\subset\text{spin}(4)$
of the R-symmetry group
\cite{Cheung:1998te,Cheung:1998wj,Ganor:2000un}. So the effective
four-dimensional theory will include the fields of the
$\N=2^*$ theory ($\N=2$ plus massive adjoint hypermultiplet) plus all the
higher Kaluza-Klein modes on the torus.

The low-energy effective
action of the four-dimensional theory will be described by
Seiberg-Witten theory and in particular we focus on the Seiberg-Witten
curve. We show how it may
be obtained by using a generalization of Witten's brane configuration
in Type IIA/M-theory \cite{Witten:1997sc}.
The resulting curve is identical, as is expected, to that proposed in
\cite{Ganor:2000un} related to
the moduli space of instantons on a non-commutative 4-torus.
The curve has also been determined in \cite{HIV} via a number of
different methods: a Dijkgraaf-Vafa
matrix model, geometric engineering and the instanton calculus.

Once we have established the form of the curve, we turn to the
question of whether there is a related integrable system. The
motivation is as follows. If we consider the $\N=2^*$ theory
strictly in four dimensions then the Seiberg-Witten curve is the
spectral curve of the complexified $N$-body elliptic
Calogero-Moser system, where the position coordinates take values
on an auxiliary torus whose complex structure is $\tau$, the
complexified coupling of the four-dimensional theory, and whose
momenta are valued in ${\bf C}$. If we then lift this theory to
five dimensions compactified on a circle, then the Seiberg-Witten
curve becomes the spectral curve of the $N$-body elliptic
Ruijsenaars-Schneider integrable system \cite{Nek,BMM1,BMM2}. This
is sometimes described as the ``relativistic" Calogero-Moser model
since the momenta are now periodic in one direction on ${\bf
C}$---as one would expect for a rapidity. The ``speed of light''
is proportional to the inverse of size of the circle so that when
the radius goes to zero the Calogero-Moser system is recovered. Now
imagine lifting the five-dimensional theory to six dimensions
compactified on a torus. The question then is: what is the next
object in the chain?
\begin{align*}
\text{Calogero-Moser}&\quad\longrightarrow&\text{Ruijsenaars-Schneider}&\quad\longrightarrow
&\quad \text{?}\qquad\qquad\\
q\in{\bf T}^2\ ,\quad p\in{\bf C}&&q\in{\bf T}^2\ ,\quad p\in{\bf
R}\times{\bf S}^1&&q\in{\bf
  T}^2\ ,\quad p\overset{\text{?}}\in{\bf T}^2\end{align*}
One naturally expects that the momenta now take values on a torus,
as indicated above. In fact the natural guess is the torus dual to
the compactification torus. Hence, with both the positions and the
momenta being doubly-periodic the hypothetical integrable system
is sometimes called the ``doubly-elliptic'', or Dell, system. Such
a Dell system, in the case of 2 particles, has been defined and
investigated in \cite{BMM3,Fock:1999ae,Braden:2001yc}. It is one
of main aims of this paper to find the hypothetical integrable
system for any number of particles. Our approach makes prominent a
polarized abelian surface with moduli $\tau$, $\rho$ and $m$, and
the system we uncover provides a concrete realization of a
Beauville-Mukai system based on the abelian surface, rather than
the more often discussed K3 surface. We will compare our system
with Dell system in due course. We also note that the curve of the
6D theory with $\N=(1,0)$ supersymmetry having $N_F=2N$ fundamental  
hypermultiplets compactified on a torus was constructed some time ago
\cite{Gorsky:1997mw}. 

\section{The Curve from M Theory}

There are several ways to find the form of the Seiberg-Witten curve
that describes compactified six-dimensional gauge theories. In the first
place, the form of the curve follows from the work of
\cite{Ganor:2000un}. In this
work, it was shown how the curve, which we denote $\Sigma$,
plays a r\^ole in the
description of the moduli space of $N$ instantons in a $U(1)$ gauge
theory on a non-commutative torus.\footnote{The case of $N$ instantons
  in a $U(k)$ gauge theory is relevant to a quiver generalization of the
  $\N=2^*$ theory where the gauge group becomes $U(N)^k$; this will be
  described elsewhere.}
Another way to engineer the curve
is via a Dijkgraaf-Vafa matrix model, although in this case, the model is
actually a two-dimensional matrix field theory defined on the
compactification torus \cite{HIV}. The matrix model describes a deformation of
the four-dimensional theory from $\N=2$ to $\N=1$. If the deformation
is suitably generic then varying it allows the vacuum to track across the
Coulomb branch of the $\N=2$ theory. In this way, the Seiberg-Witten
curve itself can be extracted.
Yet another way to engineer the curve is to use equivariant
localization techniques, pioneered in \cite{Nekrasov:2002qd,Nekrasov:2003rj},
to calculate the instanton partition function
from which the Seiberg-Witten curve can be extracted from a
saddle-point analysis. Going from the four-dimensional
theory to the compactified six dimensional one involves replacing the
integral over instanton moduli space by the partition function of a
two-dimensional sigma model whose target is the same space.
The details are described in \cite{HIV}.

In this section, we take a different approach
by proposing a form for the curve based on Witten's M-theory
construction of $\N=2$ theories \cite{Witten:1997sc}.
The answer agrees with other methods.
We start with the Type IIA configuration which describes the
four-dimensional $\N=2^*$ theory.
There is one NS5-brane whose world-volume spans
$\{x^0,x^1,x^2,x^3,x^4,x^5\}$ and $N$ D4-branes spanning
$\{x^0,x^1,x^2,x^3,x^6\}$. The direction $x^6$ is periodic.
The position of each D4-brane in $x^{4,5}$
may be described by the complex variable
$x=x^4+ix^5$. In order to incorporate the mass $m$ the spacetime must
be modified: as one goes around the $x^6$ circle
$x$ shifts by $m$.

As explained in \cite{Witten:1997sc}, after lifting to M-theory the
configuration is described by a single M5-brane whose world-volume
fills $\{x^0,x^1,x^2,x^3\}$ while the remaining directions are
described by a Riemann surface $\Sigma$ embedded in the
four-dimensional space $\Q$ whose coordinates are
$\{x^4,x^5,x^6,x^{10}\}$. The M-theory direction $x^{10}$ together
with $x^6$ are valued on a torus of complex structure $\tau$, the
coupling constant of the four-dimensional theory. We will
introduce the holomorphic coordinate \EQ{ z=\frac1{2\pi
R_{10}}\big(x^{10}+ix^6\big)\ , } where $2\pi R_{10}$ is the
circumference of the M-theory circle, and define \EQ{ {\bf
T}^2_z=\big\{z\in{\bf C}\,\big|\ z\thicksim z+p+\tau q\ ,\quad
p,q\in{\bf Z}\big\}\ . } In order to incorporate the mass $m$, the
complex $x$-plane is then fibred over this base torus so that as
one goes around the $B$-cycle of ${\bf T}_z^2$ there is a shift
$x\to x+m$. This means that, although $\Q$ is equal to ${\bf
R}^2\times{\bf T}_z^2$ locally, this is not true globally.

The curve $\Sigma\hookrightarrow\Q$ describing the geometry of the
M5-brane is the Seiberg-Witten curve of the four-dimensional
theory. The curve can be described algebraically via \EQ{
F(z,x)=x^N+f_1(z)x^{N-1}+f_2(z)x^{N-2}+\cdots+f_N(z)=\prod_{i=1}^N(x-
x_i(z))=0\ .
\label{cur4d}
}
The coefficients are fixed by the following
conditions. First, in order to incorporate the mass $m$, we need
\EQ{ F(z+1,x)= F(z,x)\ ,\qquad F(z+\tau,x)= F(z,x+m)\ . } The
eigenvalues $x_i(z)$ define a branched $N$-fold cover of the base
torus ${\bf T}_z^2$. There is a distinguished point, which we
choose at $z=0$, that corresponds to the position of the
NS5-brane. At $z=0$ on exactly one of the sheets, say $i$, a point
on $\Sigma$ which we denote as $P_0$, the associated function
$x_i(z)$ has a simple pole with residue $m$. Note that
we can define the meromorphic
function $v$ on $\Sigma$ which takes the form\footnote{In the
following $\zeta(z)$ is the Weierstrass
  $\zeta$-function with half periods $\tfrac12$ and $\tfrac\tau2$.}
\EQ{
v_j(z)=Nx_j(z)-m\big(\zeta(z)-2z\zeta(\tfrac12)\big)
}
on the $j^{\rm th}$ sheet. Although $v$ is single-valued on $\Sigma$
it now has simple poles on each sheet at $z=0$, with residues $-m$,
on sheets $j\neq i$, and with residue $(N-1)m$ on the $i^{\rm th}$
sheet. The genus of $\Sigma$ follows from the
Riemann-Hurwitz Theorem and is found to be $N$. In addition, the
number of moduli of $\Sigma$ can be computed using the Riemann-Roch
Theorem; one finds $N$.

We can view the curve $\Sigma$ as $N$ copies of the base torus
${\bf T}^2_z$ plumbed together by $N-1$ branch cuts. The $A$ and
$B$ cycles on the base torus ${\bf T}^2_z$ lift to a basis of
1-cycles $A_i$ and $B_i$, $i=1,\ldots,N$, on $\Sigma$. Let
$\omega_i$ be the associated abelian differentials of the first
kind. We can then identify $z$ as a multi-valued function on
$\Sigma$ in the following way. First of all, $z$ must shift by 1
around each of the $A_i$ cycles; this means \EQ{
z(P)=\int_{P_0}^P\sum_{j=1}^N\omega_j } where $P\in\Sigma$ and
$P_0$ is some fixed origin. Around each $B_i$ cycle $z$ must shift
by $\tau$, the coupling. Hence, we have the following constraint
on the period matrix $\Pi$ of $\Sigma$:
\EQ{
\sum_{j=1}^N\Pi_{ij}=\tau\qquad\forall i\ .
\label{pmc}
}
Another way to view the surface is in terms of the multi-valued function
$x$. This is continuous around the $A_j$ cycles but jumps by $m$
around the $B_j$ cycles. The function $x$ covers the complex plane
with the point at infinity corresponding to the distinguished
point $P_0$, the position of the NS5-brane. There are $N$ pairs of
cuts ${\cal C}_j^\pm$ with (complex)
end-points $(a_j,b_j)$ and $(a_j+m,b_j+m)$,
$j=1,\ldots,N$. The cuts in a given pair are glued together so
that the top/bottom of the upper cut is glued to the bottom/top of
the lower cut. Thus each pair generates a handle as illustrated in
Fig.~1. 
\begin{figure}
\begin{center}
\includegraphics[scale=0.5]{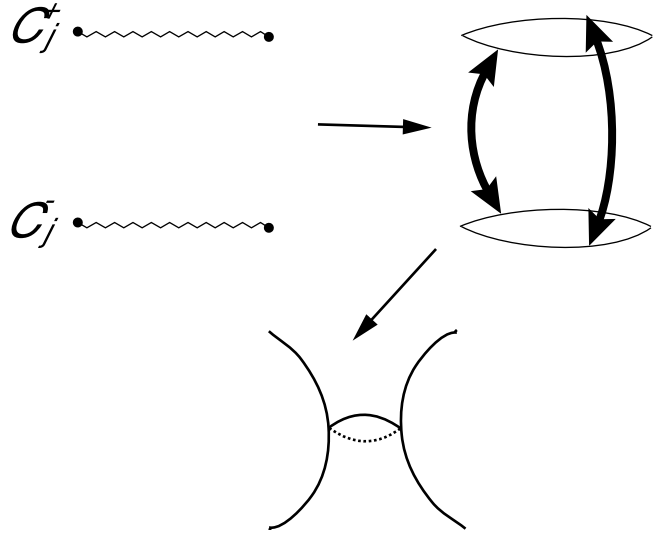}
\end{center}
\caption{\small Each pair of cuts ${\cal C}_j^-=(a_j,b_j)$ and
${\cal C}_j^+=(a_j+m,b_j+m)$ are
  identified in such a way as to form a handle on the $x$-plane.}
\end{figure}
The moduli
correspond to positions of the ends of the lower cuts, so number $2N$;
however, there are $N$ conditions \eqref{pmc} on the period matrix
so, all-in-all, there are $N$ (complex) moduli.

Now we consider the six-dimensional theory compactified on the
torus ${\bf T}^2$ defined as follows, \EQ{ {\bf
T}^2=\big\{y\in{\bf C}\,\big|\ y\thicksim
y+\frac{\beta}{\rho-\bar\rho} (p+\rho q)\ ,\quad p,q\in{\bf
Z}\big\}\ . } In order to describe the compactified
six-dimensional gauge theory in the M-theory set-up we have to
compactify $\{x^4,x^5\}$ on the dual to the compactification torus
defined as \EQ{ \tilde{\bf T}^2=\big\{y\in{\bf C}\,\big|\
y\thicksim y+\frac{2\pi i}\beta(p+\rho q)\ ,\quad p,q\in{\bf
Z}\big\}. \label{torii} } So the space $\Q$ in which $\Sigma$ is
embedded becomes a 4-torus. In order to incorporate the mass $m$,
it cannot be simply the product $\tilde{\bf T}^2\times{\bf
T}^2_z$. Rather as one goes around the $B$-cycle of the ${\bf
T}_z^2$, $x$ must shift by $m$, as in the four-dimensional
theory. This shift must be mirrored in the opposite direction.
This means that $\Q$ must be a ``slanted'' product of the two
tori. We now argue that the form of $\Q$ is determined by the
condition that there exists a curve $\Sigma$ embedded in it with
the appropriate properties. As a real manifold we can think of
$\Q$ as ${\bf R}^4/\Lambda$, where $\Lambda$ is a rank 4 lattice.
If a set of basis vectors for $\Lambda$ are $\lambda_\alpha$,
$\alpha=1,\ldots,4$, then we can think of $\Q$ as the region \EQ{
\sum_{\alpha=1}^4y_\alpha\lambda_\alpha\in{\bf R}^4 }
parameterized by four real variables $0\leq y_\alpha\leq1$ whose
faces $y_\alpha=0$ and $y_\alpha=1$ are identified. We propose
that a curve $\Sigma$ of the right form exists when $\Q$ is an
abelian surface (a 2-dimensional abelian variety) \cite{GH,LB}. To start, we
view ${\bf R}^4$ as a complex manifold ${\bf C}^2$ and introduce
holomorphic coordinates $(z_1,z_2)$. These are related to the real
coordinates $y_\alpha$ via the $2\times4$ period matrix \EQ{
z_i=\sum_{\alpha=1}^4\Omega_{i\alpha}y_\alpha\ . } In order to be
an abelian variety, the condition that it can be embedded in
projective space, $\Omega$ must satisfy certain conditions (the
Riemann conditions). With a suitable choice of basis for $\Lambda$
it can be shown that $\Omega$ may be put in the form \EQ{
\Omega=\begin{pmatrix} \delta_1 & 0 & \Gamma_{11} & \Gamma_{12}\\
0 & \delta_2 & \Gamma_{21} & \Gamma_{22} \end{pmatrix},
}
where $\delta_{1,2}$ are integers and $\delta_1|\,\delta_2$. Then the
conditions that $\Q$ be an abelian variety are that the $2\times2$
matrix $\Gamma$ is symmetric and ${\rm Im}\,\Gamma$ is positive
definite. In this basis, $(\delta_1,\delta_2)$ defines the
polarization of $\Q$. The conditions on $\Gamma$ ensures that
\EQ{
\omega=\delta_1\,dy_1\wedge dy_3+\delta_2\,dy_2\wedge dy_4\
,\label{symp}
}
is a $(1,1)$-form. The fact that $\Q$ is an
abelian surface will ensure that there exists a holomorphic curve
$\Sigma$ with the right properties. This curve arises in the
following way. A polarized abelian variety has an associated line
bundle ${\cal L}$ whose $1^{\rm st}$ Chern class $c_1({\cal
L})=[\omega]$. This line bundle admits $\delta_1\delta_2$
holomorphic sections which are constructed in terms of generalized
theta functions. Our conventions for these latter functions are
\EQ{ \Theta\left[{\delta\atop\epsilon}\right](Z|\,\Pi)
=\sum_{m\in{\bf Z}^g}\exp\big(\pi
  i(m+\delta)\cdot\Pi\cdot(m+\delta)
+2\pi i(Z+\epsilon)\cdot(m+\delta)\big)\ . } Here $Z$, $\delta$,
$\epsilon$ and $m$ are $g$-vectors and $\Pi$ is a $g\times g$
matrix (with ${\rm Im}\,\Pi$ positive definite). The basis for the
holomorphic sections of ${\cal L}$ are then given by the
$\delta_1\delta_2$ theta functions ($g=2$ in the case of an
abelian surface) \EQ{
\Theta\left[\begin{matrix}\tfrac i{\delta_1} &\tfrac j{\delta_2}\\
0&0\end{matrix}\right](z_1\ z_2|\Gamma)\quad 0\leq
i<\delta_1,\quad0\leq j<\delta_2\ . } A general section is then a
linear combination of the above. The divisor of a given section
naturally defines a curve $\Sigma$ in the homology class dual to
$c_1({\cal L})=[\omega]$. In other words, for some coefficients
$A_{ij}$, the curve $\Sigma$ is defined by \EQ{
\sum_{i=0}^{\delta_1-1}
\sum_{j=0}^{\delta_2-1}A_{ij}\Theta\left[\begin{matrix}\tfrac
i{\delta_1}
&\tfrac j{\delta_2}\\
0&0\end{matrix}\right]\big(z_1\ z_2\big|\Gamma\big)=0\ .
\label{curvb}
}

In order to realize the structure above in the M-theory scenario
we must relate the coordinates $(z_1,z_2)$ to the physical
coordinates $(z,x)$. Notice that if $\Gamma_{12}=\Gamma_{21}=0$
then $\Q$ is the product of two elliptic curves. This will
describe the degenerate limit when the mass $m=0$ and in this
limit the two elliptic curves are identified with ${\bf T}^2_z$
and $\tilde{\bf T}^2$. In other words, in this limit, we identify
the fundamental domain of ${\bf T}^2_z$ with $z=y_1+y_3\tau$ and
that of $\tilde{\bf T}^2$ with $x=(2\pi i/\beta)(y_2+y_4\rho)$. The
next fact we need is that the curve $\Sigma$ we are after must be
wrapped $N$ times around the $z$-torus but only once around the
$x$-torus. This requires some explanation. Suppose $\beta$ is small so
that the dual torus $\tilde{\bf T}^2$ is very large. In this limit
it makes sense to think of the curve as approximately equal to the
four-dimensional one. The four-dimensional curve goes off to
infinity in the $x$-plane in the neighbourhood of the
distinguished point $P_0$. In the compactified theory this region
must wrap the torus once. The multiple wrapping situation will
occur in the quiver theory generalization of our set-up and
will be described elsewhere.

So more precisely, the homology class of $\Sigma$ should be
Poincar\'e dual to the form $\omega$ in \eqref{symp} with
$\delta_1=1$ and $\delta_2=N$. In the limit $m=0$, this fixes \EQ{
\Omega\big|_{m=0}=\begin{pmatrix} 1 & 0 & \tau & 0\\
0 & N & 0 & N\rho \end{pmatrix}\ .
}
A non-zero mass $m$ corresponds to a non-vanishing off-diagonal
component $\Gamma_{12}$. Given that $x$ must shift by $m$ as one goes
around the $B$ cycle of the $z$-torus, determines
the period matrix $\Gamma$ for non-vanishing mass:
\EQ{
\Gamma=\begin{pmatrix} \tau & \tfrac{N\beta m}{2\pi i}\\ \tfrac{N\beta
  m}{2\pi i} & N\rho
\end{pmatrix}
\label{qpm}
}
and in addition we discover
\EQ{
z_1=z\ ,\qquad
z_2=\frac{N\beta x}{2\pi i}\ .
}
With this period matrix, the curve $\Sigma$ in
\eqref{curvb} has genus $N+1$. The coefficients $A_{0j}$, along
with a choice of origin for $Z$, are the moduli of the curve.
Since the $A_{0j}$ are defined up to an overall complex re-scaling
they parameterize a copy of $\P^{N-1}$. Hence the moduli
space of the curve is $\ms=\Q\times\P^{N-1}$, which is $N+1$
complex dimensional.

We can picture the curve in two ways. First, as in the
four-dimensional case, as the $x$-plane, which is now a torus with
periods $(2\pi i/\beta)(1,\rho)$, together with $N$ pairs of cuts
across which $x$ jumps by $m$ and whose edges are identified to
create a handle, as in Fig.~1. This is illustrated in Fig.~2.
\begin{figure}
\begin{center}
\includegraphics[scale=0.5]{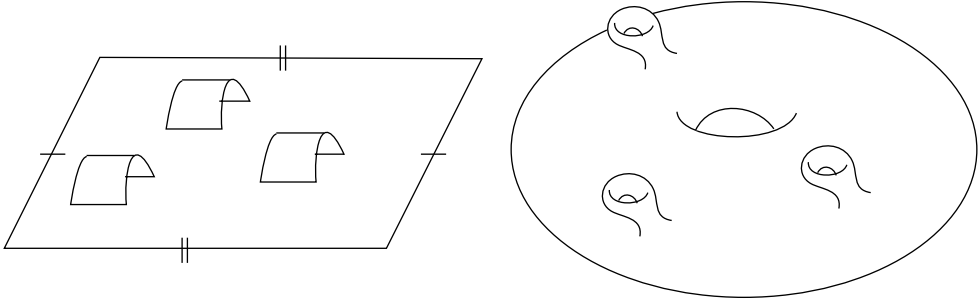}
\end{center}
\caption{\small On the left, the
surface $\Sigma$ realized as the cut $x$-torus $\tilde
{\bf T}^2$. The cuts in each of the $N$ pairs are separated by $m$ and are
glued together as in Fig.~1. On the right, an impression of the
surface realized as $N$ handles $\tilde{\bf T}^2$.}
\end{figure}
The second representation consists of $N$ copies of a torus in the
$z$-plane, with periods $(1,\tau)$, and which are joined by $N-1$
branch cuts. On the face of it, such a surface would have genus
$N$ but on one of the sheets there is a pair of cuts across which
$z$ jumps by $N\beta m/(2\pi i)$ whose edges are identified to create an extra
handle.  This is illustrated in Fig.~3.
\begin{figure}
\begin{center}
\includegraphics[scale=0.5]{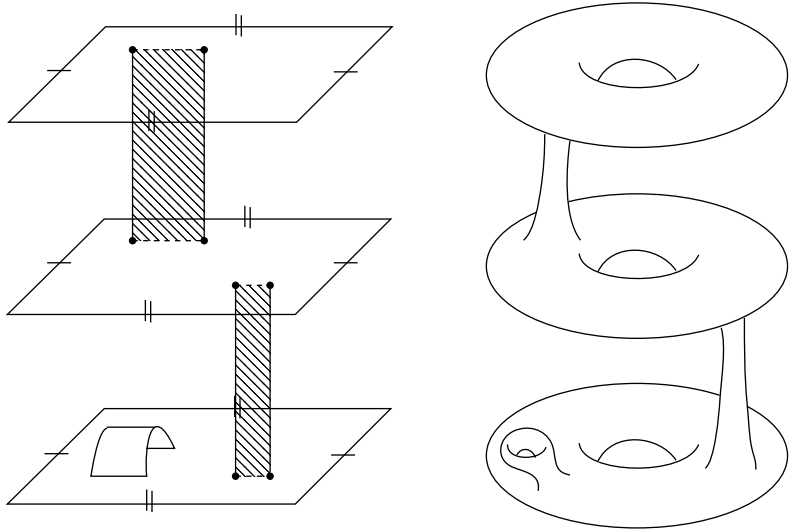}
\end{center}
\caption{\small On the left, the surface $\Sigma$ realized as $N$
copies of the $z$-torus ${\bf T}^2_z$ connected by $N-1$ branch cuts. On
one of the sheets there is an additional pairs of cuts separated by
$N\beta m/(2\pi i)$
which are glued together as in Fig.~1. On the right, is
an impression of the surface illustrating the $N$ copies of the $z$-torus
plumbed together along with the additional handle on one of the sheets.}
\end{figure}

Using the explicit expression for the theta-function, the form of
the curve \eqref{curvb} can be re-cast in the following way which
makes the reduction to five and four dimensions more immediate:
\EQ{ \sum_{n=0}^\infty\frac1{n!}\Big(\frac{m}{2\pi i}\Big)^n
\partial_z^n\theta_1(\pi z|\tau)
\partial_x^n H(x)=0\ .
\label{curva}
}
where $\theta_1$ is the usual Jacobi theta function related to
$\Theta$ for $g=1$ via
\EQ{
\theta_1(\pi z|\tau)=-\Theta\left[\begin{matrix}\tfrac12\\
\tfrac12\end{matrix}\right](z|\tau)\ . } In the above, \EQ{
H(x)=\prod_{j=1}^N\theta_1(\tfrac{\beta}{2i}(x-\zeta_j)|\rho)\ . }
Here, $\zeta_j$ are $N$ of the moduli and the remaining one
corresponds to shifting $z$ by a constant. By equating
\eqref{curva} and \eqref{curvb}, up to some constant shifts in $x$
and $z$, we deduce that \EQ{
A_{0j}(\zeta_i)=\sum_{\{i_k\}}\Theta\left[\begin{matrix}
\tfrac{i_1}N&\cdots&\tfrac{i_{N-1}}N\\
0&\cdots&0\end{matrix}\right] \big(\tfrac{N\beta}{2\pi i}
(\zeta_2-\bar\zeta)\ \cdots\
\tfrac{N\beta}{2\pi i}(\zeta_N-\bar\zeta) \big|\tilde\Pi\big)\ , } where the
$j$-dependence arises through the definition of the sum which is
over $i_k=0,\ldots,N-1$ subject to
\EQ{ j+\sum_{k=1}^{N-1}i_k\in
N\cdot{\bf Z} } and $\tilde\Pi$ is an
$(N-1)\times(N-1)$-dimensional matrix with elements \EQ{
\tilde\Pi_{ij}=\rho N(N\delta_{ij}-1)\ . } Finally \EQ{
\bar\zeta=\frac1N\sum_{i=1}^N\zeta_i\ . }

To go from the curve of the six-dimensional theory that of the
five-dimensional theory, one takes
$\rho\to i\infty$ in which case \EQ{
H(x)\to\prod_{j=1}^N\sinh(\tfrac{\beta}{2}(x-\zeta_i))\ ,
}
and from the five
to the four-dimensional theory one takes $\beta\to0$ giving rise to
\EQ{
H(x)\to\prod_{j=1}^N(x-\zeta_i)\ .
\label{h4d}
}
The curve of the
four-dimension theory is identical to the curve described by
Donagi and Witten \cite{Donagi:1995cf}. It is well-known that this
is the spectral curve of the $N$-body elliptic Calogero-Moser
integrable system \cite{Martinec:1995qn,GM96,D'Hoker:1997ha}. The
curve of the five-dimensional theory can be shown to be the
spectral curve of the Ruijsenaars-Schneider integrable system
\cite{BMM2} as predicted by Nekrasov \cite{Nek}.
The weak and strong coupling limits of these theories have been
investigated in \cite{BM01}.

Both $x$ and $z$ are multi-valued functions on the curve while
$dx$ and $dz$ are holomorphic differentials (abelian differentials
of the $1^{\rm st}$ kind) which we can identify, in terms of the basis
$\{\omega_i\}$, as
\EQ{
dx=\frac{2\pi i}\beta\omega_{N+1}\ ,\qquad dz=\sum_{i=1}^N\omega_i\ ,
\label{diffsQ}
}
with respect to a homology basis $\{A_i,B_i\}$. Here, the $A_i$
cycles, $i=1,\ldots,N$, encircle the cuts $[a_i+m,b_i+m]$ on the
$x$-torus and where $B_i$, $i=1,\ldots,N$, join the bottom cut to the
top cut in each pair (and hence because of the gluing condition are
cycles). This leaves $A_{N+1}$ and $B_{N+1}$ which are the usual $A$
and $B$
cycles on the $x$-torus $\tilde T^2$. The cycles are shown in Fig.~4.
\begin{figure}
\begin{center}
\includegraphics[scale=0.6]{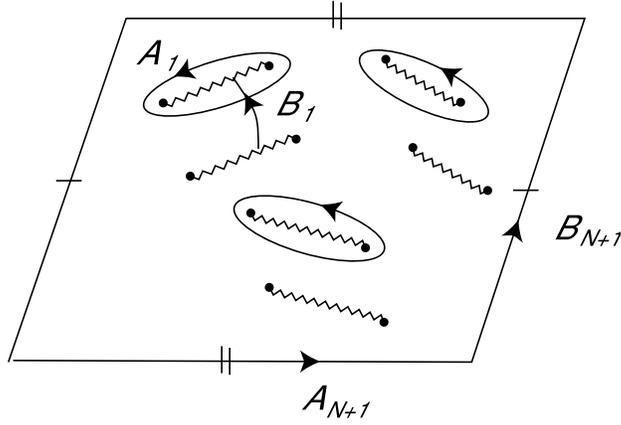}
\end{center} \caption{\small The surface $\Sigma$ realized as the
cut torus $\tilde{\bf T}^2$ showing our choice of homology basis.}
\end{figure}
The abelian differentials of the
$1^{\rm st}$ kind are then normalized by
$\oint_{A_i}\omega_j=\delta_{ij}$ and 
given this $\oint_{B_i}\omega_j=\Pi_{ij}$.
The fact that $z$ is valued on the torus ${\bf T}_z^2$ means that,
as in \eqref{pmc}, there is a condition on the period matrix of
$\Sigma$:
\EQ{ \sum_{j=1}^N\Pi_{ij}=\tau\qquad\forall i\ . }
In
addition, since $x$ jumps by $m$ around the cycles $B_i$,
$i=1,\ldots,N$, and by $2 \pi i\rho/\beta$ around the remaining cycle
$B_{N+1}$, the full period matrix of $\Sigma$ has the form \EQ{
\Pi=\begin{pmatrix} \Pi_{11}&\cdots&\Pi_{1N}& \tfrac{\beta m}{2\pi i}\\
\vdots&\ddots&\vdots & \vdots\\ \Pi_{N1}&\cdots&\Pi_{NN}&
\tfrac{\beta m}{2\pi i}\\ \tfrac{\beta m}{2\pi i}&\cdots&\tfrac{\beta m}{2\pi i}&\rho
\end{pmatrix}\ ,\quad\sum_{j=1}^N\Pi_{ij}=\tau\ .
\label{psw} } Notice that $z$ jumps by \EQ{
\oint_{B_{N+1}}\sum_{i=1}^N\omega_i=\frac{N\beta m}{2\pi i}\ , } around $B_{N+1}$ a
fact that is consistent with the period matrix of $\Q$ in
\eqref{qpm}.

There is yet another way to write the curve. First of all, one
introduces another genus $N+1$ curve $\hat\Sigma$ with a period matrix
\EQ{
\hat\Pi=\begin{pmatrix}\tau&\tfrac{\beta m}{2\pi i}&\tfrac{\beta m}{2\pi i}
&\cdots&\tfrac{\beta m}{2\pi i}\\
\tfrac{\beta m}{2\pi i}&\rho&0&\cdots&0\\
\tfrac{\beta m}{2\pi i}&0&\rho&\ddots&\vdots\\
\vdots & \vdots& \ddots & \ddots & 0 \\
\tfrac{\beta m}{2\pi i}& 0 & \cdots & 0 &\rho\end{pmatrix}\ . }
This curve is an
$N$-fold unbranched cover of a genus 2 curve. It is then not too
difficult to show that \eqref{curva} is equivalent to \EQ{
\Theta\left[\begin{matrix}\frac1{2}&\cdots&\frac1{2}\\
    \frac1{2}&\cdots&\frac1{2}\end{matrix}
\right]\big(z,\, \tfrac{N\beta}{2\pi
  i}(x-\zeta_1),\,\cdots,\,\tfrac{N\beta}{2\pi i}(x-\zeta_N)
\big|\hat\Pi\big)=0 \label{curvc}} up to constants shifts in $x$
and $z$. The expression \eqref{curvb} arises when reducing the
theta function of $\hat\Sigma$ in terms of the genus two curve and
the associated Prym variety. We see that the curve $\Sigma$ lies
in the theta divisor of $\hat\Sigma$. This means that there are
$N$ points $\hat P_i$ on $\hat\Sigma$ such that \EQ{
z=\sum_{i=1}^{N} \int^{\hat P_i}_{\hat P_0}\hat\omega_0+\hat{\cal
K}_0\ ,\quad \tfrac{N\beta}{2\pi i}
(x-\zeta_j)=\sum_{i=1}^{N} \int^{\hat P_i}_{\hat
P_0}\hat\omega_j+\hat{\cal K}_j\quad j=1,\ldots,N\ , }
where we have absorbed constant shifts in $z$ and $x$ and ${\cal K}_j$
is the vector of Riemann constants. So given
$x-\zeta_i$, $i=1,\ldots,N$, the last $N$ equations
determine the $\hat P_i$ which then
determines $z$ via the first equation.

\section{An Integrable System}

It is well-known that Seiberg-Witten curves are the spectral
curves of integrable systems. For example, the curve of the
four-dimensional theory is the spectral curve of the $N$-body
elliptic Calogero-Moser system. This is more than a happy
coincidence. In order to see this, it is helpful to take the
theory in four dimensions, in this case the $\N=2^*$ theory, and
compactify it to three dimensions on a circle of finite radius
\cite{SWred}. If the radius is very large compared with the scale
of symmetry breaking, then it is appropriate first of all to
consider the low-energy effective theory in four dimensions, in
this case a $U(1)^N$ gauge theory with gauge couplings given by
the period matrix of the Seiberg-Witten curve, \EQ{ {\cal
L}_\text{eff}=\sum_{i,j=1}^N\Big[{\rm Im}(\Pi_{ij})
F_i\wedge{}^*F_j+i{\rm Re}(\Pi_{ij}) F_i\wedge F_j\Big]\cdots\ , }
compactified on a circle to three dimensions. The
degrees-of-freedom in three dimensions consist of the Wilson lines
of the $U(1)^N$ gauge field around the circle along with the
three-dimensional gauge fields which may be dualized into scalars.
The Wilson lines and dual photons naturally combine into $N$
scalar fields which take values on a complex torus with period
matrix determined by the couplings $\Pi_{ij}$; in other words the
effective degrees-of-freedom are described by a point in the
Jacobian ${\EuScript J}(\Sigma)$ of $\Sigma$. In addition to this,
we also have the moduli of the curve itself. So, all-in-all, the
moduli space of vacua of the three-dimensional theory consists of
a fibration of ${\EuScript J}(\Sigma)$ over $\ms(\Sigma)$. This is
actually the phase space of a complexified integrable system,
where the fibre corresponds to the angle variables $\{\theta_i\}$ and
base space is parameterized by the conjugate action variables
$\{a_i\}$. As yet we don't have an explicit expression for the action
variables but this will emerge. In this context $\Sigma$, which is
fixed under evolution in the integrable system, is called the
spectral curve. The reason why the compactification to three
dimensions is useful is because the symplectic form (which in the
holomorphic context is a $(2,0)$-form) is independent of the
compactification radius and this independence means that various
holomorphic quantities that can be calculated in the
three-dimensional theory are actually valid in the
four-dimensional theory. In particular, we have in mind the vacuum
structure after breaking to $\N=1$ \cite{Dorey:1999sj,Dorey:2001qj,Hollowood:2003ds}. The above connection
with integrable systems is part of a more general setting. The
moduli space of the Seiberg-Witten curve, parameterized by the action
variables of the integrable system, is a special
K\"ahler manifold. Indeed, special K\"ahler geometries arise as
moduli spaces  in many examples of interest (for example, the
scalars of ${\cal N}=2$ four-dimensional Einstein-Maxwell
super-gravities \cite{dWvP}, or appropriate CFT's \cite{Sei88,S},
or $CY_3$'s \cite{FS89}). Freed \cite{F} has argued that the
cotangent bundle to such, with appropriate restrictions, is the
phase space of (an algebraically) completely integrable system.

In the above discussion the integrable system appears directly in
terms of action/angle variables and one can ask the question as to
whether there is a representation in terms of more familiar
dynamical variables such as the positions and momenta of
particles. It is at this point that the realization of the curve
in the physical problem to hand becomes important. The point is
that in the four-dimensional theory
the curve naturally appears as holomorphically embedded in a
four-dimensional space $\Q$ which we recall for our
four-dimensional theory is (in terms of real geometry) locally of
the form ${\bf T}^2_z\times
{\bf R}^2$. We shall soon see that this
embedding provides the position and momentum basis. The curve
\eqref{cur4d} (or equivalently \eqref{curva} with $H(x)$ as in
\eqref{h4d}) has $N+1$ moduli
corresponding to the $\zeta_i$, $i=1,\ldots,N$, along with the
choice of origin for $z$. The origin for $z$ is fixed by choosing
the position of the NS5-brane to be $z=0$ as in Section 2. Note
that the choice of origin for $x$ is incorporated in the average
$\sum_{i=1}^N\zeta_i/N$; we choose $\sum_{i=1}^N\zeta_i=0$. This
leaves $N-1$ of the moduli to vary giving a subspace
$\ms_0\subset\ms(\Sigma)$. These moduli are parameterized by $N-1$
action variables which are the canonically conjugate variables to
an $N-1$ dimensional subspace of the Jacobian ${\EuScript
J}_0\subset {\EuScript J}(\Sigma)$ defined by the condition
$\sum_{i=1}^N\theta_i=0$. (Recall that in the four-dimensional
case the curve $\Sigma$ has genus $N$.)

We now show that the remaining non-trivial part of the phase space
consisting of $N-1$ action and $N-1$ angle variables are
bi-rationally equivalent to the degrees-of-freedom of $N-1$ points
in the ambient space $\Q$ with local coordinates $Z_i=(q_i,p_i)$
in ${\bf T}^2_z\times{\bf R}^2$. The equivalence is
straightforward to describe. Generically at least, there is a
unique curve $\Sigma$ in the reduced moduli space (of dimension
$N-1$) which goes through $N-1$ points in $\Q$. This curve
naturally has $N-1$ marked points on it $P_j$, $j=1,\ldots,N-1$.
These points map to a point in the reduced Jacobian $\J_0$ by first
mapping into the Jacobian $\J(\Sigma)$
via the Abel map:
\EQ{
\theta_i=\sum_{j=1}^{N-1}\int_{P_0}^{P_j}
\omega_i\ , }
where $P_0$ is a fixed base point, and then by using
the projection $\J(\Sigma)\to\J_0$. It is
straightforward to follow the map in the other direction. The fact
that the equivalence is only bi-rational refers to the fact that
the map may break down at certain non-generic points and in
particular when some of the points $(q_i,p_i)$ coincide.

One can go on to show that the bi-rational equivalence between the
positions and momenta and the angle and
action variables (where the latter are as yet unspecified)
is a canonical transformation. We will do this
for the six-dimensional case later. The resulting description of
the integrable system in terms of $N-1$ points in $\Q$ is exactly
what one expects for the $N$-body elliptic Calogero-Moser system.
In this case, the $N-1$ quantities $q_i$ and $p_i$ represent the
relative positions of the $N$ particles. So the projection onto
$\ms_0$ and ${\EuScript J}_0$ simply removes the trivial
centre-of-mass motion.

The above picture works in the same way for the five-dimensional
theory. In this case, $\Q$ is locally $ {\bf S}^1\times{\bf
R}\times{\bf T}^2_z$ and, in particular, the ``momenta'' are now
valued on ${\bf S}^1\times{\bf
  R}$, as a real manifold. This is appropriate for a relativistic
system where the $\beta p_i$ play the r\^ole of the rapidities. It is
indeed one way of describing the Ruijsenaars-Schneider system, as
the relativistic generalization of Calogero-Moser. Once again, the
reduced system describes the relative motion of an $N$-body
system.

We now turn to the six-dimensional theory. One's first thought is
that the six-dimensional theory will require the momenta to take
values on a torus, in fact the dual torus $\tilde{\bf
  T}^2$. So with the positions taking values on the torus ${\bf
  T}^2_z$, this leads one to expect the existence of the so-called
``doubly elliptic'', or Dell, integrable system
\cite{BMM3,Fock:1999ae,Braden:2001yc}. However, we note that the
mass parameter $m$ evident in our description of the ambient space
$\Q$ means that in general we are not dealing simply with a global
product of two tori.\footnote{We note that for suitable rational
values of $\beta m/2\pi i$ we note that $\Q$ does however take on
this product structure.} Thus the system we construct is more than
``doubly elliptic''. We shall compare our construction with that
of Dell system shortly. 

The main proposition of this paper is that even in the
six-dimensional theory, $\Sigma$ is the spectral curve of a
completely integrable mechanical system, albeit of a rather
unusual kind. In particular, the phase space of the system can be
viewed, as in the four-dimensional case, as the moduli space of
the curve $\Sigma$, with the Jacobian ${\EuScript J}(\Sigma)$
fibered over it. In this section, we show that the integrable
system can be thought of as describing the interactions of a set
of ``particles'' whose whose positions and momenta $(q,p)$, as a
2-vector, take values that are the local coordinates of a point on
the abelian surface $\Q$. The fact that the momenta take values in
a compact space means that the analogy with a set of particles is
not to be taken too literally. However, once the $x$-torus
decompactifies in one, or both, directions, so the curve reduces
to the five and four-dimensional one, respectively, then the
system really can be interpreted in terms of a system of
particles.

Much of following is identical to the four-dimensional case, but
there are some differences.
Loosely speaking the phase space of the integrable system is
identified with the fibre bundle whose base is the moduli space of
the curve $\Sigma$ and whose fibre is the Jacobian of the curve
$\J(\Sigma)$. However, in the six-dimensional case the separation
of the ``centre-of-mass'' factors is more involved. Recall that
the moduli space of the curve, $\ms(\Sigma)$, and hence the base
of the fibration, can be identified as the product \EQ{
\ms(\Sigma)\simeq\Q\times\P^{N-1} } which has (complex) dimension
$N+1$. The total space of the bundle consequently has the form
$\Q\times\X$, where $\X$ is a fibering of $\J(\Sigma)$ over
$\P^{N-1}$. From the embedding of $\Sigma\hookrightarrow \Q$ we
may pull back the differentials $dz$ and $dx$ on $\Q$ to abelian
differentials of the $1^{\rm st}$ kind on $\Sigma$ (\ref{diffsQ}).
This gives us a natural embedding of the abelian surface
$\iota:\Q\hookrightarrow{\EuScript J}(\Sigma)$. This
embedding means that there is an abelian
subvariety ${\EuScript J}_0$ of ${\EuScript J}(\Sigma)$ such that
$\Q\oplus{\EuScript J}_0$ is isogenous to
${\EuScript J}(\Sigma)$. Hence, $\X$ can be viewed as a fibered
product of the fixed $\Q$ and $\Y$, where $\Y$ is a complex $2N-2$
dimensional space. The space $\Y$ itself a non-trivial fibering of the abelian
subvariety ${\EuScript J}_0\subset{\EuScript J}(\Sigma)$ over the
reduced moduli space $\ms_0=\P^{N-1}$.

Before delving into
the technicalities, the idea that will emerge is that the space $\X$ is
bi-rationally equivalent to the symmetric product $\text{Sym}^N(\Q)$
\cite{Vafa:1995zh,Vafa:1995bm,Bershadsky:1995qy} (see also
\cite{Gorsky:1999rb}). We can equivalently
think of this as the Hilbert scheme $\text{Hilb}\sp{[N]}\Q$. If we take local
coodinates $(q,p)=(z,x)$ on $\Q$, then the $N$ points $(q_i,p_i)$
define the positions and ``momenta'' of $N$ particles. Separating out
the centre-of-mass motion, leaves us with the $N-1$ relative positions
and momenta and this corresponds to the reduced
phase space that we denoted $\Y$ above.

Now we will identify more explicitly the reduced phase $\Y$ and show
that it is bi-rationally equivalent to the centred motion of $N$
points in $\Q$. Let us denoted the coordinates in
${\EuScript J}(\Sigma)$
by $\theta_i$, $i=1,\ldots,N+1$. We may describe the
embedding of the abelian surface $\iota(\Q)\hookrightarrow
{\EuScript J}(\Sigma)$ concretely as follows:
\EQ{
\iota(z,x)=\big(z,z,\ldots,z,\tfrac{N\beta}{2\pi i}x\big)\ .
}
From our discussion above,
the subspace ${\EuScript J}_0$ consists of the orthogonal
subspace to this embedding ${\EuScript J}(\Sigma)=\Q\oplus\J_0$.
The phase space of the reduced system $\Y$, the fibration of
${\EuScript J}_0$ over $\P^{N-1}$, is then bi-rationally equivalent
to the space of $N-1$ points in $\Q$ (the relative coordinates of $N$
particles) with local coordinates
$Z_i=(q_i,p_i)$. The relation between the
action/angle variables of the reduced system $\{a_i\}$ and $\{\theta_i\}$
and the $N-1$ points in $\Q$ mirrors the
discussion above for the four-dimensional case. First of all, we fix
the overall position of $\Sigma$ in $\Q$. Then, at least generically, given
$N-1$ points in $\Q$ there exists a unique curve $\Sigma\subset\Q$
containing these points. If $P_j$, $j=1,\ldots,N-1$, are these
points on $\Sigma$ then we may obtain a point in the
reduced Jacobian $\J_0$ by first mapping into the
Jacobian $\J(\Sigma)$ via the Abel map:
\EQ{
\theta_i=\sum_{j=1}^{N-1}\int_{P_0}^{P_j}\omega_i\ ,
\label{Abelm}
}
where $P_0$
is a fixed base point, and then by using the projection
$\J(\Sigma)\to\J_0$. Under this
equivalence, the symplectic form is preserved so
that\footnote{Using $\delta$ rather than $d$ for differential
forms on the phase space, avoids confusion with differential forms
on $\Sigma$.}
\EQ{ \sum_{i=1}^{N+1}\delta a_i\wedge\delta\theta_i=
\sum_{i=1}^{N-1}\delta p_i\wedge\delta q_i\ .
\label{symf}
}
Here, the one-forms $\{\delta\theta_i\}$ are constrained to lie in the
reduced Jacobian $\J_0$ so that the number of independent
degrees-of-freedom match on both sides. The
right-hand side of \eqref{symf} just corresponds to $N-1$ copies
of the standard symplectic form on $\Q$.

We now prove \eqref{symf}. Suppose we choose $z$ to be a local
coordinate on $\Sigma$. The curve would be described by the
holomorphic function $x=x(z)$. Consider the abelian differentials
of the first kind on $\Sigma$, $\omega_i$, $i=1,\ldots,N+1$.
Explicitly in this coordinate system $\omega_i=f_i(z)dz$ and
\eqref{Abelm} becomes
\EQ{
\theta_i=\sum_{j=1}^{N-1}
\int^{q_j}_{q_0}f_i(z)dz\ .
}
Now suppose we vary the $Z_i$. The point in the Jacobian will vary as
\EQ{ \delta\theta_i=\sum_{j=1}^{N-1}f_i(q_j)\,\delta q_j\ .
\label{era} }
Now we turn to the action variables. How do we
describe these? As we vary the actions we change our curve
$\Sigma$ and this will be described by the normal bundle
$N_\Sigma=T\Q |_\Sigma/T\Sigma$ to $\Sigma$ in $\Q$. Then,
using the holomorphic symplectic form, $T\Q\cong T\sp\star\Q$ and
consequently $N_\Sigma\cong T\sp\star\Sigma$. Thus we may identify
our actions with holomorphic sections of $T\sp\star\Sigma$, \EQ{
x(z)\,dz=\sum_{j=1}^{N+1} a_j\,\omega_j(z)\ . } Using the contour
integrals around the 1-cycles $A_j$ associated to the basis
$\omega_j$: \EQ{ a_j=\oint_{A_j}x\,dz\ , } where on the right-hand
side a pull-back from $\Q$ to $\Sigma$ is implied. Then $x\,dz$
plays the role of the Seiberg-Witten differential. Hence under a
variation at the point with coordinate $z$ \EQ{
\sum_{j=1}^{N+1}\delta a_j\, \omega_j(z)=\delta x(z)\,dz\ . } Now
we evaluate this at $z=q_i$ where $\delta x(q_i)=\delta p_i$,
giving \EQ{ \delta p_i=\sum_{j=1}^{N+1}\delta a_jf_j(q_i)\ .
\label{erb} } Using first \eqref{erb} and then \eqref{era}, the
result \eqref{symf} follows: \EQ{ \sum_{i=1}^{N-1}\delta
p_i\wedge\delta q_i= \sum_{i=1}^{N-1}\sum_{j=1}^{N+1}
\big(f_j(q_i)\delta a_j\big)\wedge\delta q_i=
\sum_{i=1}^{N-1}\sum_{j=1}^{N+1}\delta
a_j\wedge\big(f_j(q_i)\delta q_i\big) =\sum_{j=1}^{N+1}\delta
a_j\wedge\delta\theta_j\ . } Note that the right-hand side only
depends on $N-1$ independent variations $\{\delta\theta_j\}$.

\subsection{Example: the $N=2$ system}

We will describe how this picture arises in the case of $N=2$.
This case is rather simple because the integrable system
describing the relative motion is only one-dimensional. Here our
curve $\Sigma$ has genus 3 and we have that ${\EuScript
J}(\Sigma)$ is isogenous to $\Q\oplus{\EuScript J}_0$
where now ${\EuScript J}_0$ corresponds to an elliptic curve. Thus
our reduced phase space $\Y$ consists of an elliptic curve fibred over
$\ms_0\equiv\P^{1}$. This is in fact a Kummer K3 surface, which
alternately appears as the fibre $\pi\sp{-1}(0)$ of the projection
$\pi:\,\text{Hilb}\sp{[2]}\Q
\rightarrow\text{Sym}^2(\Q)\rightarrow\Q$ which comes from using
the group structure on $\Q$.

 Thus for the $N=2$ theory,
$\ms_0$ is described by a single modulus which we parameterize by
$H=-A_{00}/A_{01}$. In this case, we can map the reduced system to
a single point in $\Q$ with local coordinate $(q,p)$. According to
the discussion in the last section the curve whose overall
position in $\Q$ is determined by requiring that it goes through
the point $(q,p)$. It follows trivially that \EQ{
H(q,p)=\frac{\Theta\left[\begin{matrix}0&\tfrac 12\\
0&0\end{matrix}\right]\left(q\quad
\tfrac{\beta}{\pi i}p\,\Big|\begin{matrix}  \tau& \tfrac{\beta m}{\pi i}\\
\tfrac{\beta m}{\pi i}&2\rho\end{matrix}\right)}
{\Theta\left[\begin{matrix}0&0\\
0&0\end{matrix}\right]\left(q\quad
\tfrac{\beta}{\pi i}p\,\Big|\begin{matrix}  \tau& \tfrac{\beta m}{\pi i}\\
\tfrac{\beta m}{\pi i}&2\rho\end{matrix}\right)}\ .
}
We now show that this Hamiltonian reduces to the Hamiltonian of the
centre-of-mass motion of the
2-body elliptic Ruijsenaars-Schneider system and then the elliptic
Calogero-Moser system on decompactification of $\tilde{\bf T}^2$
in one and then two directions, respectively.

Let us take the five-dimensional limit by taking the limit
$\rho\to i\infty$. In this limit, up to a simple re-scaling,
we have
\EQ{
H_\text{5d}(q,p)=e^{\beta p}\frac{\Theta\left[\begin{matrix}0\\
    0\end{matrix}\right]\big(q+\tfrac{\beta m}{2\pi i}\big|\tau\big)}
{\Theta\left[\begin{matrix}0\\
    0\end{matrix}\right]\big(q\big|\tau\big)}+
e^{-\beta p}\frac{\Theta\left[\begin{matrix}0\\
    0\end{matrix}\right]\big(q-\tfrac{\beta m}{2\pi i}\big|\tau\big)}{
\Theta\left[\begin{matrix}0\\
    0\end{matrix}\right]\big(q\big|\tau\big)}\ .
}
In order to reproduce the conventional Hamiltonian of the
centre-of-mass motion of the 2-body elliptic
Ruijsenaars-Schneider system, we must shift
\EQ{
p\to p-\frac1{2\beta}\log\frac{\Theta\left[\begin{matrix}0\\
    0\end{matrix}\right]\big(q+\tfrac{\beta m}{2\pi i}\big|\tau\big)}{
\Theta\left[\begin{matrix}0\\
    0\end{matrix}\right]\big(q-\tfrac{\beta m}{2\pi i}\big|\tau\big)}\
,\qquad
q\to q+\tfrac12(1+\tau)\ .
}
In which case, up to another simple re-scaling,
\EQ{
H_\text{5d}(q,p)=\cosh(\beta p)\sqrt{\wp(\tfrac{\beta m}{2\pi i})
-\wp(q)}\ ,
}
where $\wp(q)$ is the Weierstrass function with periods 1 and $\tau$.

The Hamiltonian appropriate to the four-dimensional theory
is obtained by taking $\beta\to0$ which yields, again up to a
simple re-scaling,
\EQ{
H_\text{4d}(q,p)=p^2+\frac{m^2}{2\pi^2}\wp(q)\ ,
}
which is the well-known
Hamiltonian of the two-particle elliptic Calogero-Moser system.

It remains to discuss the connection between the integrable system
of this paper and that of the Dell system \cite{Braden:2001yc}. (We shall use
the notation of the latter paper when making comparison.) Both
approaches involve a genus three curve in a $(1,2)$ polarized
abelian variety: here we had $\Sigma\hookrightarrow\Q$, while for
the Dell system this was $\bar{\cal C}_z\hookrightarrow S$. Also, in both
approaches, after removing the center of mass coordinates from
${\EuScript J}(\Sigma)$ (respectively ${\EuScript J}(\bar{\cal
C}_z)$) we are left with (for the $N=2$ case) a one-dimensional
abelian variety, or elliptic curve, here ${\EuScript J}_0$ and
there $\pi\sp\star (E_z)$. (Further, both works give a
construction involving Prym varieties.) The Hamiltonians
constructed in these different approaches look rather different
and it remains to connect them more directly. This will be left
for a later work \cite{bh}. The advantage of the approach adopted in
the present paper is that it immediately generalizes to $N>2$, a
generalization that was previously unknown.

\section{Discussion}

The present paper has shown how to construct the Seiberg-Witten
curve of the six-dimensional $\N=(1,1)$ gauge theory compactified
on a torus to four dimensions. The argument given was based on
Witten's M-theory construction of $\N=2$ theories but the curve we
arrive may be naturally constructed from various approaches
\cite{Ganor:2000un,HIV}. Central to our discussion was the
appearance of a $(1,N)$ polarized abelian variety $\Q$ specified
by the complex structures $\tau$, $\rho$ and mass $m$. In addition
to the four-dimensional coordinates of the M5-brane, this was
specified by a holomorphic curve $\Sigma$ embedded in $\Q$, of
genus $N+1$: the Seiberg-Witten curve. We presented this curve
explicitly in various forms via theta functions showing agreement
with the curve of \cite{Ganor:2000un} describing $N$ instantons in
a $U(1)$ theory on a non-commutative torus. The quiver
generalization to a $U(N)^k$ gauge theory, associated to $N$
instantons in $U(k)$, will simply change the
polarization of the abelian variety under consideration.

We also have demonstrated how the curve $\Sigma$ may be viewed as
the spectral curve of an integrable system which generalizes the
$N$-body elliptic Calogero-Moser and Ruijsenaars-Schneider systems
associated to four and five-dimensional gauge theories. The
resulting system is not simply doubly elliptic as the positions
and momenta, as two-vectors, take values in the ambient space
$\Q$. The integrable system we have constructed is rather natural:
the curve $\Sigma$ in $\Q$ is described by the linear system
$\P(H\sp0(\Q,{\cal O}({\cal L}))\sp\star)\cong\P\sp{N-1}$, our
space of reduced actions, together with a translation in $\Q$, our
``centre-of-mass" coordinates. Together with our actions
${\EuScript J}(\Sigma)$ these form a $2(N+1)$ dimensional phase
space which may be understood as a particular example of a class
of integrable systems discovered by Mukai \cite{Muk84, Muk87} who
constructed the symplectic structure on the moduli space of stable
sheaves on a symplectic surface (a particular example being our
abelian surface $\Q$). If one considers the moduli space ${\cal
M}_\Q(0,c_1,c_2)$ of sheaves on $\Q$ of rank $0$, $c_1=c_1({\cal
L})$ and some $c_2$ then one gets a symplectic moduli space, which
is a relative Jacobian ({\it i.e.\/}, union of Jacobians). The
fact that the rank is zero here means these are torsion sheaves
and $c_1=c_1({\cal L})$ says that the support of the sheaves is a
curve, which is cohomologous to our curve $\Sigma$. Our phase
space is bi-rational to this moduli space which itself is
bi-rational to the product of the Hilbert scheme
$\text{Hilb}\sp{[N]}\Q$ with $\Q$. We exhibited the
canonical transformation between the action-angle variables and
the set of $N-1$ points in $\Q$ corresponding to the reduced phase
space. Finally, although we showed that our construction has many
features in common with the known Dell system for two particles,
we have yet to relate the two rather different descriptions of the
Hamiltonians. We shall return to this and other features of our
integrable system in \cite{bh}.

Another interesting issue regarding the new integrable system is
whether it admits a Lax-type representation. One would expect that such
a representation should follow from an associated Hitchin system. The
Hitchin system should follow from a D-brane construction involving the
four torus $\Q$, as in
\cite{Ganor:2000un,Vafa:1995bm,Bershadsky:1995qy,Kapustin:1998pb}.
This will be investigated elsewhere.

To conclude, the construction of our paper has been premised on a fixed
polarized Abelian variety $\Q$ specified by the complex structures
$\tau$, $\rho$ and mass $m$. In general we could consider each
such theory over the moduli space of such polarized Abelian
varieties. The moduli space of polarized Abelian varieties is
itself well studied \cite{HKW,GP} and we observe that there is a
natural $sl(2,\Z)\times sl(2,\Z)$ acting on
this. It is interesting to ask how string theory sweeps out this
moduli space.

TJH would like to thank the organizers of the Simons Workshop on
Mathematics and Physics for providing an excellent environment for
developing some of these ideas. HWB would like to thank Eyal
Markman and Tom Bridgeland for helpful remarks.


\begin{thebibliography}{99}

{\small

\bibitem{SW94}
N.~Seiberg and E.~Witten, ``Electro-magnetic duality, monopole
condensation and confinement in ${\cal N}=2$ supersymmetric
Yang-Mills theory," Nucl.Phys., {\bf B426} (1994) 19-53, {\tt
[arXiv:hep-th/9407087]}.

\bibitem{Cheung:1998te}
Y.~K.~Cheung, O.~J.~Ganor and M.~Krogh,
``On the twisted (2,0) and little-string theories,''
Nucl.\ Phys.\ B {\bf 536} (1998) 175
{\tt[arXiv:hep-th/9805045]}.

\bibitem{Cheung:1998wj}
Y.~K.~Cheung, O.~J.~Ganor, M.~Krogh and A.~Y.~Mikhailov,
``Instantons on a non-commutative T(4) from twisted (2,0) and  little-string theories,''
Nucl.\ Phys.\ B {\bf 564} (2000) 259
{\tt[arXiv:hep-th/9812172]}.

\bibitem{Ganor:2000un}
O.~J.~Ganor, A.~Y.~Mikhailov and N.~Saulina,
``Constructions of non commutative instantons on T(4) and K(3),''
Nucl.\ Phys.\ B {\bf 591} (2000) 547
{\tt[arXiv:hep-th/0007236]}.


\bibitem{Witten:1997sc}
E.~Witten,
``Solutions of four-dimensional field theories via M-theory,''
Nucl.\ Phys.\ B {\bf 500} (1997) 3
{\tt[arXiv:hep-th/9703166]}.


\bibitem{HIV}
T.~J.~Hollowood, A.~Iqbal and C.~Vafa,
``Matrix Models, Geometric Engineering and Elliptic Genera,''
{\tt arXiv:hep-th/0310272}.


\bibitem{Nek}
N.~Nekrasov,
``Five dimensional gauge theories and relativistic integrable systems,''
Nucl.\ Phys.\ B {\bf 531} (1998) 323
{\tt[arXiv:hep-th/9609219]}.


\bibitem{BMM1}H. W. Braden, A. Marshakov, A. Mironov and A. Morozov,
``Seiberg-Witten Theory for a Nontrivial Compactification from
Five-Dimensions to Four-Dimensions", Phys.\ Lett.\ B {\bf  448}
(1999) 195-202  {\tt[arXiv:hep-th/9812078]}.

\bibitem{BMM2}H. W. Braden, A. Marshakov, A. Mironov and A. Morozov, ``The
Ruijsenaars-Schneider Model in the Context of Seiberg-Witten
Theory", Nucl.\ Phys.\ B{\bf 558} (1999) 371-390
{\tt[arXiv:hep-th/9902205]}.


\bibitem{BMM3}H. W. Braden, A. Marshakov, A. Mironov and A. Morozov, ``On
Double-Elliptic Integrable Systems. 1. A Duality Argument for the
case of SU(2)", Nucl.\ Phys.\ B{\bf 573} (2000) 553-572
{\tt[arXiv:hep-th/9906240]}.

\bibitem{Fock:1999ae}
V.~Fock, A.~Gorsky, N.~Nekrasov and V.~Rubtsov, ``Duality in
integrable systems and gauge theories,'' JHEP {\bf 0007} (2000)
028 {\tt[arXiv:hep-th/9906235]}.


\bibitem{Braden:2001yc}
H.~W.~Braden, A.~Gorsky, A.~Odessky and V.~Rubtsov,
``Double-elliptic dynamical systems from generalized Mukai-Sklyanin  algebras,''
Nucl.\ Phys.\ B {\bf 633} (2002) 414
{\tt[arXiv:hep-th/0111066]}.

\bibitem{Gorsky:1997mw}
A.~Gorsky, S.~Gukov and A.~Mironov,
``SUSY field theories, integrable systems and their stringy/brane  origin. II,''
Nucl.\ Phys.\ B {\bf 518} (1998) 689
{\tt[arXiv:hep-th/9710239]}.


\bibitem{Nekrasov:2002qd}
N.~A.~Nekrasov,
``Seiberg-Witten prepotential from instanton counting,''
{\tt arXiv:hep-th/0206161}.


\bibitem{Nekrasov:2003rj}
N.~Nekrasov and A.~Okounkov,
``Seiberg-Witten theory and random partitions,''
{\tt arXiv:hep-th/0306238}.


\bibitem{GH}
P.~Griffiths and J.~Harris, ``Principles of Algebraic Geometry,'' Wiley 1978.


\bibitem{LB}
H.~Lange and Ch.~Birkenhake, ``Complex Abelian Varieties,''
Springer-Verlag 1992.



\bibitem{Donagi:1995cf}
R.~Donagi and E.~Witten,
``Supersymmetric Yang-Mills Theory And Integrable Systems,''
Nucl.\ Phys.\ B {\bf 460} (1996) 299
{\tt[arXiv:hep-th/9510101]}.

\bibitem{Martinec:1995qn}
E.~J.~Martinec,
``Integrable Structures in Supersymmetric Gauge and String Theory,''
Phys.\ Lett.\ B {\bf 367} (1996) 91
{\tt[arXiv:hep-th/9510204]}.


\bibitem{GM96}
A.~Gorsky, A.~Marshakov, ``Towards effective topological gauge
theories on spectral curves,'' Phys.\ Lett. B {\bf375} (1996)
127-134 {\tt[arXiv:hep-th/9510224]}.


\bibitem{D'Hoker:1997ha}
E.~D'Hoker and D.~H.~Phong,
``Calogero-Moser systems in SU(N) Seiberg-Witten theory,''
Nucl.\ Phys.\ B {\bf 513} (1998) 405
{\tt[arXiv:hep-th/9709053]}.


\bibitem{BM01}
H. W. Braden and A. Marshakov, ``Singular Phases of Seiberg-Witten
Integrable Systems: Weak and Strong Coupling,'' Nucl.\ Phys.\ B
{\bf 595} (2001) 417-466 {\tt[arXiv:hep-th/0009060]}.


\bibitem{SWred}
N.~Seiberg and E.~Witten, ``Gauge Dynamics And
Compactification To Three Dimensions,''
{\tt[arXiv:hep-th/9607163]}.


\bibitem{Dorey:1999sj}
N.~Dorey,
``An elliptic superpotential for softly broken N = 4 supersymmetric  Yang-Mills theory,''
JHEP {\bf 9907} (1999) 021
{\tt[arXiv:hep-th/9906011]}.


\bibitem{Dorey:2001qj}
N.~Dorey, T.~J.~Hollowood and S.~Prem Kumar,
``An exact elliptic superpotential for N = 1* deformations of finite  N = 2 gauge theories,''
Nucl.\ Phys.\ B {\bf 624} (2002) 95
{\tt[arXiv:hep-th/0108221]}.

\bibitem{Hollowood:2003ds}
T.~J.~Hollowood,
``Critical points of glueball superpotentials and equilibria of  integrable systems,''
{\tt arXiv:hep-th/0305023}.

\bibitem{dWvP}
B. de Wit and A. Van Proyen, ``Potentials and symmetries of
general gauged supergravity-Yang-Mills models," Nucl.\ Phys.\ B
{\bf 245} (1984) 89-117.

\bibitem{Sei88}
N. Seiberg, ``Observations on the moduli space of superconformal
field theories'', Nucl.\ Phys.\ B {\bf 303} (1988) 286.

\bibitem{S}
A. Strominger, ``Special geometry,'' Commun. Math. Phys. {\bf 133}
(1990) 163-180.

\bibitem{FS89}
S. Ferrara and A. Strominger, ``N=2 Sspace-time supersymmetry and
Calabi-Yau moduli space'',  in \emph{Strings '89}, eds. R.
Arnowitt, R. Bryan, M.J. Duff, D.V. Nanopoulos and C.J. Pope
(World Scientific, Singapore, 1989) p245.


\bibitem{F}
Daniel S.~Freed, ``Special K\"ahler manifolds,'' Commun.Math.Phys.
{\bf 203} (1999) 31-52 {\tt[arXiv:hep-th/9712042]}.

\bibitem{Vafa:1995zh}
C.~Vafa,
``Gas of D-Branes and Hagedorn Density of BPS States,''
Nucl.\ Phys.\ B {\bf 463} (1996) 415
{\tt[arXiv:hep-th/9511088]}.


\bibitem{Vafa:1995bm}
C.~Vafa,
``Instantons on D-branes,''
Nucl.\ Phys.\ B {\bf 463} (1996) 435
{\tt[arXiv:hep-th/9512078]}.

\bibitem{Bershadsky:1995qy}
M.~Bershadsky, C.~Vafa and V.~Sadov,
``D-Branes and Topological Field Theories,''
Nucl.\ Phys.\ B {\bf 463} (1996) 420
{\tt[arXiv:hep-th/9511222]}.



\bibitem{Gorsky:1999rb}
A.~Gorsky, N.~Nekrasov and V.~Rubtsov,
``Hilbert schemes, separated variables, and D-branes,''
Commun.\ Math.\ Phys.\  {\bf 222} (2001) 299
{\tt[arXiv:hep-th/9901089]}.

\bibitem{bh}
H.~W.~Braden and T.~J.~Hollowood, {\it to appear\/}.



\bibitem{Muk84}
S.~Mukai, ``Symplectic structure of the moduli space of
sheaves on an abelian or $K3$ surface'', Invent. Math. (1984){\bf
77} 101-116.

\bibitem{Muk87}
S.~Mukai, ``Moduli of vector bundles on $K3$ surfaces and
symplectic manifolds'', (Japanese) Sugaku Expositions {\bf 1}
(1988), no. 2, 139--174; (English) Sugaku {\bf 39} (1987), no. 3,
216--235.


\bibitem{HKW}
K.~Hulek, C.~Kahn and S.~H.~Weintraub, ``Moduli
Spaces of Abelian Surfaces: Compactification, Degenerations and
Theta Functions,'' de Gruyter, Berlin 1993.


\bibitem{GP}
M.~Gross and S.~Popescu, ``Calabi-Yau Threefolds and Moduli
of Abelian Surfaces I,'' {\tt[arXiv:math.AG/0001089]}.

\bibitem{Kapustin:1998pb}
A.~Kapustin and S.~Sethi,
``The Higgs branch of impurity theories,''
Adv.\ Theor.\ Math.\ Phys.\  {\bf 2} (1998) 571
{\tt[arXiv:hep-th/9804027]}.






}\end{thebibliography}
\end{document}